\documentclass[1p]{elsarticle}
\usepackage{graphicx}
\usepackage{enumerate}
\newcommand \beq{\begin{eqnarray}}
\newcommand \eeq{\end{eqnarray}}

\newcommand{\vect}[1]{{\mathbf #1}}

\begin{document}

\title{Chiral Symmetry Breaking in Monolayer Graphene \\
 by Strong Coupling Expansion of Compact and Non-compact \\
 U(1) Lattice Gauge Theories}
\author{Yasufumi Araki}
\address{Department of Physics, The University of Tokyo, Tokyo 113-0033, Japan}

\begin{abstract}
Due to effective enhancement of the 
  Coulomb coupling strength in the vacuum-suspended graphene, 
   the system may turn from a semimetal into an insulator by the formation of 
  a gap in the fermionic spectrum.
This phenomenon is analogous to the spontaneous breaking of 
  chiral symmetry in  the strong-coupling relativistic field theories.
We study this ``chiral symmetry breaking''  
  and associated collective excitations on graphene in the strong coupling regime 
   by taking  U(1) lattice gauge theory as an effective model for graphene. 
Both compact and non-compact formulations of the U(1) 
 gauge action show chiral symmetry breaking with equal magnitude of the  
  chiral condensate (exciton condensate)
   in the strong coupling limit, while they start to deviate
  from  the next-to-leading order in the strong coupling expansion.
Phase and amplitude fluctuations of the order parameter
   are also investigated: in particular, a  mass formula for the pseudo-Nambu--Goldstone mode
    ($\pi$-exciton),  which is analogous to Gell-Mann--Oakes--Renner relation for 
     the pion in quantum chromodynamics (QCD), is
     derived from the axial Ward-Takahashi identity.
 To check the applicability of the effective field theory description, 
 typical energy scales of fermionic and bosonic excitations 
 are estimated by identifying the lattice spacing of the 
 U(1) gauge theory with that of the original honeycomb lattice of graphene.
\end{abstract}
\maketitle

\section{Introduction}

Graphene, a monoatomic layer material of carbon atoms with honeycomb lattice structure,
has attracted a great interest both in theoretical and experimental sides.
One of its important features is the ``Dirac cone'' structure:
since the charge carriers on graphene at low energy obey the linear dispersion relation
around two independent ``Dirac points'' in the 1st Brillouin zone \cite{wallace_1947},
they can be described as Dirac quasiparticles with U(4) chiral symmetry,
which means that their bare mass $m=0$ \cite{Semenoff:1984dq}.
Thus graphene shows a gapless spectrum and the density of states vanishes at zero-energy,
which means that monolayer graphene behaves as a 2-dimensional semimetallic material.
It exhibits many interesting phenomena:
high carrier mobility, anomalous quantum Hall effect, and so on \cite{castro_neto_2009}.

There is one critical difference between the quasiparticles on graphene and ordinary relativistic fermions:
the Fermi velocity $v_{_F}$ of the quasiparticles on graphene,
which is determined by physical quantities characteristic to the material,
is about 300 times smaller than the speed of light $c$.
On the other hand, the propagation speed of the electromagnetic field still remains $c$
in the vacuum.
Such discrepancy between two velocities leads to an effective enhancement of the Coulomb coupling strength,
suggesting the possibility that the chiral symmetry breaking (exciton condensation)
would spontaneously occur and that the spectrum would be gapped in the vacuum-suspended graphene \cite{CN09}.
Since the effective coupling strength is larger than that in QED about 300 times,
perturbative approach is unsuitable for this system.
Thus, we need various non-perturbative treatments,
which is analogous to the situation in quantum chromodynamics (QCD) \cite{Hatsuda:1994pi}.
Although experimental investigation of vacuum-suspended graphene started only recently \cite{Bolotin_2008},
this problem is becoming important in the industrial aspect as well as in the theoretical side.

There have been various theoretical studies about this problem.
Studies with the gap equation (Schwinger--Dyson equation) based on $1/r$ Coulomb potential,
first performed in 2001,
have found that graphene would be gapped in the strong-coupling and small-flavor region.
They have also estimated critical values of the coupling strength $\alpha$ and the flavor number $N$
\cite{khveshchenko_2001,gorbar_2002,gorbar_2009}.
The infrared behavior of the long-range Coulomb interaction is investigated by exact renormalization group methods
and critical exponents of the phase transition are estimated \cite{Giuliani_2010}.
The above studies are focused on the critical behavior of the system
 around the critical values $\alpha_{_C}$ and $N_C$.

The behavior of the system in the large-$N$ limit has also been studied by $1/N$ expansion.
It has been seen that the long-range Coulomb interaction becomes irrelevant
and that the spectrum would not be gapped at large $N$,
even though the coupling strength is large enough \cite{Herbut_2006,son_2007,son_2008}.

Recently lattice Monte Carlo simulations have been performed
based on the Thirring-like 4-fermi interaction \cite{hands_2008}
and on the U(1) lattice gauge theory \cite{drut_2009,giedt_2009,drut_2010}.
They have calculated chiral condensate (the order parameter for chiral symmetry breaking)
with varying coupling strength and have estimated the critical coupling strength,
with compact and non-compact gauge formulations.
They have also estimated critical exponents of the phase transition and the equation of state in the symmetry-broken region.
Due to the finite size effect of the lattice,
the above lattice Monte Carlo simulations are limited to the finite coupling strength ($\alpha < \infty$)
and finite bare mass ($m > 0$) region.
The behavior of the system in the chiral limit ($m=0$),
such as the critical coupling value,
is estimated by extrapolation with the help of the equation of state,
introduced in analogy with (3+1)-dimensional QED.

In this paper, we investigate the behavior of this system
in/around the strong coupling ($\alpha \rightarrow \infty$) and chiral ($m =0$) limit analytically
by strong coupling expansion of the square-lattice-regularized effective gauge theory model.
This method is one of the suitable methods to investigate the non-perturbative behavior of gauge theories,
such as QCD \cite{Drouffe:1983fv,Nishida:2003uj,Miura:2009nu},
and has been first applied to the graphene effective model in our previous paper \cite{Araki_Hatsuda_2010}.
U(1) lattice gauge theory is employed, as a low-energy effective model
for monolayer graphene with Coulomb interaction at zero temperature.
An expansion around the strong coupling and the chiral limit is performed,
which is referred to as ``strong coupling expansion'' of lattice gauge theory.
Then we calculate the value of the exciton condensate
up to the next-to-leading order (NLO) in the strong coupling expansion,
and show the spontaneous chiral symmetry breaking in/around the strong coupling limit.
Results from the compact and non-compact formulations of the gauge action are compared.
These results coincide in the strong coupling limit due to the suppression of the pure gauge term,
and deviate from the NLO in the strong coupling expansion.
Our analytic results are also compared
with the results from numerical simulations in Ref.\cite{drut_2010}.

We also investigate two fluctuation modes (bosonic collective excitations) of the exciton condensate,
which have not been studied in the preceding literatures.
One of them, which we refer to as ``$\pi$-exciton,''
behaves as a pseudo-Nambu--Goldstone boson
emerging from the spontaneous symmetry breaking.
A mass formula for the $\pi$-exciton is derived from axial Ward--Takahashi identity,
which is analogous to the Gell-Mann--Oakes--Renner (GMOR) relation
for pions in QCD \cite{gmor}.
The mass of the other one, the ``$\sigma$-exciton,'' is also calculated,
up to the NLO in the strong coupling expansion.
In order to check the applicability of the Dirac fermion description for graphene at low energy,
we compare the excitation energies (masses) of the above fermionic and bosonic excitations
to the typical energy scale from the interatomic spacing of the original honeycomb lattice of graphene.



This paper is organized as follows:
in Section \ref{sec:model},
we briefly review the continuum effective model and square lattice model
for monolayer graphene.
In Section \ref{sec:strong-coupling},
we perform strong coupling expansion with the lattice effective model
and derive the free energy of this system up to the NLO,
both with the compact and non-compact formulations of the gauge action.
We calculate the value of the chiral condensate from this effective potential.
In Section \ref{sec:col-ex},
$\pi$- and $\sigma$-exciton modes are investigated.
Masses of these modes are calculated,
and a GMOR-like mass formula
is derived.
Finally in Section \ref{sec:conclusion},
we summarize our work and present several problems for future studies.

\section{Low-energy effective model}
\label{sec:model}

In this section, we briefly review the effective field theory
which can well describe the low-energy electronic behavior of monolayer graphene.
We start from the tight-binding Hamiltonian \cite{wallace_1947}
\begin{equation}
H= -t \sum_{\vect{r} \in A} \sum_{i=1,2,3} \left[ a^\dag(\vect{r}) b(\vect{r}+\vect{s}_i) + b^\dag(\vect{r}+\vect{s}_i) a(\vect{r}) \right]. \label{eq:tight-binding}
\end{equation}
Here, $a^\dag,a$ and $b^\dag,b$ denote creation and annihilation operators of electrons
on the triangular (Bravais) sublattices A and B respectively,
which constitute the honeycomb lattice of graphene.
$\vect{r}$ takes the positions on the A sublattice.
$\vect{s}_i \, (i=1,2,3)$ denote the relative position of one B-site from its neighboring A-site,
and $t$ corresponds to the hopping amplitude between two neighboring sites.
This Hamiltonian has vanishing eigenvalues at two independent points $\vect{K}_\pm$,
which are called ``Dirac points,'' in the momentum space \cite{wallace_1947}.
Since the dispersion relation can be linearized around the Dirac points,
electron/hole excitations on graphene at low energy can be described as
massless Dirac fermions \cite{Semenoff:1984dq}.
Since there are 8 degrees of freedom
corresponding to 2 (number of sublattices) $\times$ 2 (number of Dirac points)
$\times$ 2 (up and down spin),
the quasiparticles on monolayer graphene can be formulated by
two ``flavors'' of four-component Dirac fermions.
Thus we can construct ``Dirac spinors'' around two Dirac points $\vect{K}_\pm$ as
\begin{equation}
\psi_\sigma(\vect{p}) =
\left( \begin{array}{cc}
a(\vect{K}_+ +\vect{p}) \\
b(\vect{K}_+ +\vect{p}) \\
b(\vect{K}_- +\vect{p}) \\
a(\vect{K}_- +\vect{p})
\end{array} \right).
\label{eq:Dirac}
\end{equation}
Here we take $|\vect{p}|$ much smaller than $|\vect{K}_\pm|$,
which we refer to as ``low-energy approximation.''
$\sigma=\uparrow,\downarrow$ denotes the original spin of the quasiparticle,
which is treated as ``flavor'' degree of freedom of the 4-component spinor.


\subsection{Effective model in continuum limit}

In order to incorporate the Coulomb interaction between the quasiparticles,
we add the electromagnetic field, namely U(1) gauge field,
in the effective action.
The model is described by so-called ``braneworld'' or
``reduced QED''\cite{gorbar_2001}-like model action \cite{gorbar_2002,son_2007}
\begin{equation}
S_E = \sum_{f}
 \int dx^{(3)} \ \bar{\psi}_f \left( D[A] +m \right) \psi_f + \frac{1}{4g^2} \sum_{\mu,\nu=1,2,3,4}  \int dx^{(4)} (\partial_\mu A_\nu - \partial_\nu A_\mu)^2 ,
 \label{eq:effaction}
\end{equation}
in the Euclidean space-time,
where the natural unit ($\hbar=c=1$) is taken.
In this model, the fermion quasiparticles run in the (2+1)-dimensional plane
$x^{(3)}=\left(\tau (=x_4),x_1,x_2 \right)$ with the Fermi velocity $v_{_F}$,
while the U(1) gauge field propagates in the (3+1)-dimensional
space $x^{(4)}=\left(\tau (=x_4),x_1,x_2,x_3 \right)$ with the speed of light $c(=1)$.
The Fermi velocity is defined by two parameters characteristic to graphene lattice,
 $v_{_F}=(3/2)t a_{_{\rm Hc}}=3.02\times 10^{-3}$, with the interatomic spacing of the honeycomb lattice
  $a_{_{\rm Hc}}=|\vect{s}_i|=1.42\ \mathrm{\AA}$ and the hopping amplitude $t \simeq 2.8\ \mathrm{eV}$ \cite{reich}.
The above value of $v_{_F}$ is the physical value observed on $\mathrm{SiO_2}$ substrate.
The ``flavor'' $f$ of the 4-component Dirac spinor $\psi_f$
runs from 1 to the ``number of flavors'' $N$.
In this paper we specifically focus on $N=2$ case,
which corresponds to the spin degree of freedom of the monolayer graphene.

The Dirac operator in Eq.(\ref{eq:effaction}) is defined as
\begin{equation}
D[A]= \gamma_4(\partial_4+iA_4) + v_{_F} \sum_{i=1,2} \gamma_i (\partial_i +i A_i) , \label{eq:DA}
\end{equation}
where $A_\mu$ denotes $\mu$-th component of the U(1) gauge field.
Since the fermions are confined in the layer,
$z$-component does not appear in the Dirac operator.
The gauge coupling constant
for the vacuum-suspended graphene $g^2=e^2/\epsilon_0$, with $e$ being the electric charge and 
 $\epsilon_0$ being the dielectric constant of vacuum.
If the layer is placed on a substrate,
interaction strength $g^2$ is screened by the factor $2/(1+\varepsilon)$,
with $\varepsilon$ being the dielectric constant of the substrate \cite{son_2007}.
Strictly speaking,  $v_F$ in Eq.(\ref{eq:DA}) is 
 a bare value of the Fermi velocity which receives a finite renormalization due to interactions.
In the strong coupling expansion which we employ,
bare value of $v_F$ is assumed to be as small as the physical value.

The Hermitian $\gamma$ matrices obey the well-known Clifford algebra 
  $\{\gamma_\mu,\gamma_\nu\}=2\delta_{\mu\nu}$,
and the ``chirality'' matrix $\gamma_5$ is defined by
$\gamma_5=\gamma_4 \gamma_1 \gamma_2 \gamma_3$.
By linearizing the tight-binding Hamiltonian in Eq.(\ref{eq:tight-binding}),
three of the gamma matrices
\begin{equation}
\gamma_1 = \left(
\begin{array}{cc}
i\tau_2 & 0 \\
0 & i\tau_2
\end{array}\right), \quad
\gamma_2 =
\left( \begin{array}{cc}
i\tau_1 & 0 \\
0 & i\tau_1
\end{array} \right), \quad
\gamma_4 =
\left( \begin{array}{cc}
\tau_3 & 0 \\
0 & -\tau_3
\end{array} \right)
\end{equation}
are uniquely obtained
corresponding to the Dirac spinor representation in Eq.(\ref{eq:Dirac}).
$\tau_i \, (i=1,2,3)$ are the Pauli matrices which acts on the sublattice (A,B) subspace.
Due to the absence of $z$-direction in the tight-binding Hamiltonian,
there is a degree of freedom in choosing a matrix representation of $\gamma_3$ and $\gamma_5$.
One possible choice is
\begin{equation}
\gamma_3 = \left(
\begin{array}{cc}
0 & -i\tau_3 \\
i\tau_3 & 0
\end{array}\right), \quad
\gamma_5 =
\left( \begin{array}{cc}
0 & \tau_3 \\
\tau_3 & 0
\end{array} \right).
\end{equation}
The bare mass in Eq.(\ref{eq:effaction}) corresponds to half of the explicit bandgap
which originates from the difference between on-site energies on A- and B-sublattices.
It is suggested that such a gap may be formed artificially on epitaxially grown graphene on substrate \cite{zhou_2007}
  or on graphene nanoribbon and nanomesh \cite{bai_2010}.
 Although it still remains a great question in graphene physics how to open a finite spectral gap,
 in this paper we leave the mass term for convenience of calculation.

\subsection{Effective coupling strength and instantaneous approximation}

To set the Dirac operator independent of the Fermi velocity $v_{_F}$,
we perform a scale transformation in the temporal direction
\begin{equation}
\tau \rightarrow \tau /v_{_F}, \quad A_4 \rightarrow v_{_F} A_4,
\end{equation}
yielding the effective mass $m_*=m/v_{_F}$.
This scale transformation renders the Dirac operator Lorentz invariant,
\begin{equation}
D_*[A]= \gamma_4(\partial_4+iA_4) + \sum_{i=1,2} \gamma_i (\partial_i +i A_i) ,
\end{equation}
and the pure gauge action (the second line in Eq.(\ref{eq:effaction}))
Lorentz non-invariant,
\begin{eqnarray}
S_G  &=& \frac{v_{_F}}{2g^2} \sum_{i=1,2,3}  \int dx^{(4)} (\partial_i A_0 - \partial_0 A_i)^2 \label{eq:effaction-e} \\
 && +\frac{1}{4g^2 v_{_F}} \sum_{i,j=1,2,3}  \int dx^{(4)} (\partial_i A_j - \partial_j A_i)^2. \label{eq:effaction-m}
\end{eqnarray}
Such Lorentz non-invariance leads to the difference between 
the coupling strength of electric (temporal) component and that of the magnetic (spatial) component
of the gauge field.
As for the magnetic component, it is weakened as $g_m^2 =v_{_F} g^2$,
rendering the coefficient of Eq.(\ref{eq:effaction-m}) sufficiently large
to apply saddle point approximation over the spatial components of the gauge field.
The saddle point gives the vanishing magnetic field, $\vect{B} = \vect{\nabla} \times \vect{A} =0$,
and the arbitrariness in choosing $\vect{A}$ can be absorbed in the local U(1) gauge invariance.
Therefore we can neglect the spatial components of the gauge field $A_i \, (i=1,2,3)$.
This approximation is usually referred to as ``instantaneous approximation,''
because it omits the retardation (magnetic) effect of the electric field.
Fluctuation correction beyond the saddle point approximation corresponds
to the weak coupling expansion by $g_m$.
In this paper, we work on the saddle point and consider only the temporal component of the gauge field.

On the other hand, the coupling strength of the electric component is enhanced as $g_*^2 =g^2/v_{_F}$,
which is about 300 times larger than the usual Coulomb coupling strength in QED.
This means that perturbative expansion by $g_*$ as for usual QED does not work well
in this graphene model,
while expansion by the inverse coupling strength
\begin{equation}
\beta = \frac{1}{g_*^2}
= \frac{\epsilon_0 (1+\varepsilon) v_{_F}}{2e^2} \label{eq:beta-g}
\end{equation}
can be well performed in this regime.
From Eq.(\ref{eq:beta-g}), we obtain the value of $\beta$ in the vacuum 0.0369,
  while that on the SiO$_2$ substrate 0.101.
Although the physical value of $\beta$ can be shifted due to the renormalization of $v_{_F}$,
here we employ the bare value of $\beta$.

\subsection{Regularization on square lattice}

In order to regularize this theory,
we introduce a square lattice
so that we can compare our results to those of the Monte Carlo simulations \cite{drut_2009,giedt_2009,drut_2010}.
We set the lattice spacing $a$ comparable to the honeycomb lattice spacing $a_{_{\rm Hc}}$
to reproduce the physical momentum cutoff on the honeycomb lattice.
Na\"{i}vely the fermionic term of Eq.(\ref{eq:effaction}) is discretized
on the square lattice with 4-component Dirac spinor $\psi$ as
\begin{equation}
S_F= \sum_{x^{(3)},y^{(3)}} \bar{\psi}_f(x) \left[D[U](x,y)+ m_* \delta_{x,y}\right] \psi_f(y), \label{eq:latticeaction-F-naive}
\end{equation}
where the lattice Dirac operator reads
\begin{equation}
D[U](x,y) = 
 \frac{1}{2} \sum_{\mu=1,2,4}
\left[\delta_{y,x+\hat{\mu}} U_\mu(x) - \delta_{x,y+\hat{\mu}} U_\mu^*(y) \right]\gamma_\mu
. \label{eq:naive-D}
\end{equation}
Here we take the lattice unit, in which all the dimensionful quantities
are scaled by $a$.
$\hat{\mu}$ denotes the unit vector in $x_\mu$-direction.
The U(1) gauge field is represented by the link variable $U_\mu(x)$
which corresponds to the link between $x$ and $x+\hat{\mu}$.
The time-like link is defined as $U_4(x)=\exp\left[i\theta(x)\right]\;(-\pi\leq\theta\leq\pi)$,
while the space-like links $U_{1,2,3}(x)$ are set to unity as a result of instantaneous approximation.
Eq.(\ref{eq:latticeaction-F-naive}) is the lattice-regularized version of the continuum action Eq.(\ref{eq:effaction})
with the finite lattice spacing $a$.
However, we can see in momentum space that such a lattice fermion has 7 unnecessary poles
other than $p_\mu=0$,
which means that the lattice action describes $2^3(=8)$ species of fermions \cite{Nielsen-Ninomiya}.
Such species are called ``doublers'', and this problem is referred to as the ``doubling problem.''

In order to avoid such doubling problem,
one possible solution is to consider the doublers as ``real'' degrees of freedom.
Since the Dirac operator Eq.(\ref{eq:naive-D}) is diagonalized by the transformation
\begin{equation}
\chi(x) \equiv \gamma_4^{x_4} \gamma_1^{x_1} \gamma_2^{x_2} \psi(x),
\end{equation}
the lattice action is written with the single-component Grassmann number $\chi(x)$.
We identify 8 doublers of $\chi$ as 4(spinor) $\times$ 2(flavor) degrees of freedom of the fermion,
which is called ``staggered fermion'' formulation \cite{Susskind_1977,Sharatchandra_1981}.
As a result, 
the lattice action for fermions on graphene reads \cite{drut_2009}
\begin{equation}
S_F= \sum_{x^{(3)}}  \left[  \frac{1}{2} \sum_{\mu=1,2,4}
 \left( V_{\mu}^+(x)-V_{\mu}^-(x) \right)  + m_{*} M(x) \right] ,
\label{eq:latticeaction-F}
\end{equation} 
with fermionic bilinears
\begin{eqnarray}
 M(x)&=& \sum_{s} \bar{\chi}_{s}(x)\chi_{s}(x),  \\
 V_{\mu}^+(x)&=& \sum_{s} \eta_{\mu}(x)\bar{\chi}_{s}(x)U_{\mu}(x) \chi_{s}(x+\hat{\mu}), \nonumber \\
 V_{\mu}^-(x)&=& \sum_{s} \eta_{\mu}(x)\bar{\chi}_{s}(x+\hat{\mu}) U_{\mu}^{\dagger}(x)   \chi_{s}(x),
\end{eqnarray}
where $\mu=1,2,4$,
and the flavor index $s$ of the staggered fermion runs from 1 to $N/2$.
In the case of the monolayer graphene,
 we need no staggered flavor index $s$, because $N/2=1$ \cite{hands_2008,drut_2009}.
The staggered phase factors  $\eta_{\mu}$ 
  corresponding to the Dirac $\gamma$-matrices are
$ \eta_4(x)=1, \eta_1(x)=(-1)^{\tau},  \eta_2(x)=(-1)^{\tau+x_1}$.
$ \epsilon(x) \equiv (-1)^{\tau+x_1+x_2}$ corresponds to the chirality $\gamma_5$.

As for the pure gauge action term,
there are two ways for discretization on the lattice.
Na\"{i}vely the discretized gauge action is written with the compact link
variables $U_4(x)$ as
\begin{equation}
S_G^{\rm (C)} = \frac{1}{g_*^2} \sum_{x^{(4)}} \sum_{j=1,2,3}
\left[1-{\rm Re} \left( U_4(x) U_4^{\dagger}(x+\hat{j}) \right)\right].
 \label{eq:latticeaction-G}  
\end{equation}
However, it is known in the normal QED that such compact gauge formulation
suffers from the so-called ``monopole condensation'' problem,
leading to anomalous phase transition \cite{monopole-cond}.
In order to avoid this problem,
we can employ the non-compact gauge formulation for this system \cite{drut_2009}:
\begin{equation}
S_G^{\rm (NC)} = \frac{1}{2g_*^2} \sum_{x^{(4)}} \sum_{j=1,2,3} \left[\theta(x)-\theta(x+\hat{j})\right]^2. \label{eq:latticeaction-G-NC}
\end{equation}
We will compare the results from these two formulations later.

\subsection{Chiral symmetry of the system}

In the chiral limit ($m\rightarrow 0$),
the continuum action Eq.(\ref{eq:effaction}) is invariant under U(4) chiral transformation
generated by 16 generators
$\{1, \vec{\sigma}\} \otimes \{1,  \gamma_3, \gamma_5, \gamma_3 \gamma_5\}$ for each flavor,
where $\sigma_i\,(i=1,2,3)$ are the Pauli matrices acting on the spin subspace.
Note that invariance under continuous chiral symmetry within the low-energy approximation
originates from the discrete $Z_2$ symmetry between
two triangular (Bravais) sublattices (A,B) of the original honeycomb lattice of graphene.
If higher order terms in momentum are introduced perturbatively,
continuous approximate symmetry is eventually broken into discrete symmetry.
Lack of $\gamma_3$ in the Dirac operator,
which comes from the fact that the fermion is confined in the $(2+1)$-dimensional layer,
extends the well-known chiral symmetry generated by $\{1, \gamma_5\}$
into such a large symmetry.
If the chiral condensate $\langle \bar{\psi}\psi \rangle$ obtains a finite expectation value,
the chiral symmetry is spontaneously broken as
\begin{equation}
\mathrm{U(4)} \rightarrow \mathrm{U(2)} \times \mathrm{U(2)}.
\end{equation}

As for the lattice-regularized model with staggered fermions (Eq.(\ref{eq:latticeaction-F})), 
the symmetry of the system is reduced into U(1)$_{_{\rm V}}$ $\times$ U(1)$_{_{\rm A}}$
 in the chiral limit:
\begin{eqnarray}
\!\!\!\!\!\!\!\!\!\! \mathrm{U(1)_V}: && \!\!\!\! (\chi(x), \bar{\chi}(x)) \rightarrow (e^{i\xi_{_{\rm V}}} \chi(x), e^{-i\xi_{_{\rm V}}}\bar{\chi}(x)) \\
\!\!\!\!\!\!\!\!\!\! \mathrm{U(1)_A}: && \!\!\!\! (\chi(x), \bar{\chi}(x)) \rightarrow (e^{i\xi_{_{\rm A}} \epsilon(x)} \chi(x), 
  e^{i\xi_{_{\rm A}} \epsilon(x)} \bar{\chi}(x) )
\end{eqnarray}
  These are remnants
   of global U(4) chiral symmetry of Eq.(\ref{eq:effaction}) \cite{Kogut:1982ds}.
Part of the chiral symmetry (flavor symmetry) is not conserved
in the staggered fermion formulation, which is known as ``taste breaking.''
Here we discuss only the spontaneous breaking of the U(1)$_{_{\rm A}}$ symmetry.
Under  the U(1)$_{_{\rm A}}$ rotation, the fermion bilinears transform as
 $M(x) \rightarrow e^{2i\xi_{_{\rm A}} \epsilon(x)}M(x)$ 
 and $V_{\mu}^{\pm}(x) \rightarrow V_{\mu}^{\pm}(x)$, so that
  the   chiral condensate $\langle \bar{\chi} \chi \rangle$ serves as an order   
  parameter for  the spontaneous symmetry breaking,
\begin{equation}
\mathrm{U(1)_V} \times \mathrm{U(1)_A} \rightarrow \mathrm{U(1)_V}.
\end{equation}
If $\langle \bar{\chi} \chi \rangle$ acquires a finite expectation value,
the global U(1)$_{_{\rm V}}$ $\times$ U(1)$_{_{\rm A}}$ symmetry is broken to
the mere U(1)$_{_{\rm V}}$ symmetry.

\section{Strong Coupling Expansion}
\label{sec:strong-coupling}

Strong coupling expansion on lattice,
which employs the inverse of the coupling constant as an expansion parameter,
has been used for analyzing the non-perturbative behavior of gauge theories \cite{Drouffe:1983fv}.
It was first applied for the pure Yang--Mills theory
to show the mechanism of confinement of quarks in the strong coupling limit \cite{Wilson_1974}.
It was also applied to SU($N$) lattice gauge theory with fermions
to show the spontaneous breaking of chiral symmetry and corresponding meson mass spectra
in the strong coupling limit successfully \cite{Kawamoto_Smit_1981}.
Recently strong coupling limit study of lattice QCD is extended to the finite temperature and density region,
to map the phase diagram of QCD matter under extreme conditions \cite{Nishida:2003uj,Nishida_2004,Kawamoto_2007}.
Finite coupling effect (NLO and next-to-NLO in strong coupling expansion) \cite{Miura:2009nu,Ohnishi_2007,TZN_2010_1}
and the confinement (Polyakov loop) effect \cite{TZN_2010_2} are also being considered by strong coupling expansion.

Since QCD is an asymptotic free gauge theory,
in which the bare coupling approaches zero in the continuum limit,
 the strong coupling expansion does not give results more than a qualitative estimate.
On the other hand, in the case of graphene,
the lattice constant $a \sim a_{_{\rm Hc}}$ is fixed to be 
a finite value and the bare coupling strength at that scale is large enough.
Therefore, the strong coupling expansion may provide us with 
not only qualitative estimate but also some quantitative calculations of physical observables.


\subsection{Link integration}

The expansion parameter for strong coupling expansion is defined in Eq.(\ref{eq:beta-g}) as $\beta \equiv 1/g_{*}^2$,
 so that the strong coupling limit corresponds to $\beta \rightarrow 0$.
 Then, the partition function can be expanded by $S_G \sim O(\beta)$
around the strong coupling limit as
\begin{equation}
Z= \int [d\chi d\bar{\chi}][d\theta] \left[ \sum_{n=0}^{\infty}
 \frac{(-S_G)^n }{n!} e^{-S_F} \right] = \int
 [d\chi d\bar{\chi}] e^{-S_{\chi}}.
 \label{eq:part-Z}
\end{equation}
Since the integrand of the above equation is written as a polynomial of link variables by the expansion,
$\theta$ integration can be analytically performed
 order by order in $\beta$ \cite{Drouffe:1983fv}.
 When the link variable $e^{i\theta}$ and its complex conjugate $e^{-i\theta}$ cancel with each other,
 the  fermion self-interaction terms are induced;
 the terms in which link variables remain vanish away
 through the link integration.
 
\begin{figure}[tbp]
\vspace{0.5cm}
\begin{center}
\includegraphics[width=7cm]{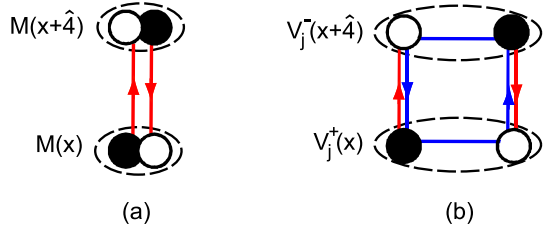}
\end{center}
\vspace{-0.5cm}
\caption{Induced four-fermion interaction in the strong coupling
 expansion.
The open and filled circles represent
 fermion fields $\chi$ and $\bar{\chi}$, respectively.
(a) In the LO,  the time-like links (red arrows) 
in the fermion action $S_F$ 
 cancel with each other to leave a spatially local interaction.
(b) In the NLO,
 the time-link links in $S_F$ are canceled by the time-like links in 
 a plaquette $S_G$  (blue arrows)  to leave a spatially non-local interaction.
}
\label{fig:links}
\end{figure}

In the LO ($O(\beta^0)$) of strong coupling expansion,
there are no difference between the compact and non-compact gauge formulations,
because the pure gauge term $S_G$ vanishes at $\beta=0$.
Thus only the fermionic term $S_F$ contributes to the LO effective action $S_\chi^{(0)}$.
Due to the Grassmann nature of $\chi$,
we can rewrite the temporal link integration part as
\begin{eqnarray}
&& \int [d\theta] \exp\left[-\frac{1}{2}\sum_{x^{(3)}} \left( V_4^+(x)-V_4^-(x) \right)\right] \\
&& \quad =
\int [d\theta] \prod_{x^{(3)}\!,\,s} \left[ 1-\frac{1}{2}\bar{\chi}_{s}(x) e^{i\theta(x)} \chi_{s}(x+\hat{4})\right]
\left[ 1+\frac{1}{2}\bar{\chi}_{s}(x+\hat{4}) e^{-i\theta(x)} \chi_{s}(x)\right]. \nonumber
\end{eqnarray}
By integrating out the U(1) compact link variables, we obtain
\begin{eqnarray}
&& \!\!\!\!\!\! \!\!\!\!\!\! \exp\left[-S_\chi^{(0)}\right] \nonumber \\
&& \!\!\!\!\!\! \!\!\!\!\!\! = \prod_{x^{(3)}} \left\{ \sum_{n=0}^{\infty} \! \left(\frac{1}{n!}\right)^{\!\! 2} \!\! \left[-\frac{1}{4} \!\! \left(\sum_{s=1}^{N/2} \bar{\chi}_{s}(x) {\chi}_{s}(x+\hat{4}) \right) \!\!\! \left(\sum_{s'=1}^{N/2} \bar{\chi}_{s'}(x+\hat{4}) {\chi}_{s'}(x) \right) \! \right]^{\! n} \! \right\} \label{eq:4-fermi-multi} \\
&& \!\!\!\!\!\! \!\!\!\!\!\! \quad\quad \times \exp\left\{ -\sum_{x^{(3)}}  \left[ \frac{1}{2} \sum_{j=1,2} 
  \left( V_j^+(x)-V_j^-(x) \right)  + m_{*} M(x) \right] \right\},
\end{eqnarray}
where the factor in the second line (Eq.(\ref{eq:4-fermi-multi}))
consists of fermion interaction up to $2N$-fermi terms.
Hereafter we fix the flavor number $N$ to 2 (monolayer case),
so that the interaction is limited to 4-fermi term:
\begin{eqnarray}
S_{\chi}^{(0)} &=& \sum_{x^{(3)}}  \left[ \frac{1}{2} \sum_{j=1,2} 
  \left( V_j^+(x)-V_j^-(x) \right)  + m_{*} M(x) \right] \label{eq:4-fermi-free} \\
  && \quad -\frac{1}{4} \sum_{x^{(3)}} M(x) M(x+\hat{4}). \label{eq:4-fermi-0}
\end{eqnarray}
Here the 4-fermi term in the second line (Eq.(\ref{eq:4-fermi-0})) comes from
the product of the temporal hopping terms $V_4^+(x)$ and $V_4^-(x)$ in $S_F$
as shown in Fig.\ref{fig:links}(a).
 
In the NLO ($O(\beta^1)$),
two gauge formulations give some difference.
With the compact formulation, the fermionic term $S_F$ and one plaquette from $S_G$ contribute
to the NLO effective action
\begin{equation}
S_{\chi}^{(1) {\rm C}} = \frac{\beta}{8} \sum_{x^{(3)}} \sum_{j=1,2}
\left[( V_j^+(x) V_j^-(x+\hat{4}) + (V_j^+ \leftrightarrow V_j^-) \right] \label{eq:4-fermi-1},
\end{equation}
which come from the product of the temporal hopping terms $V_4^+(x)$ and $V_4^-(x+\hat{j})$,
and the plaquette $U_4^\dag(x)U_4(x+\hat{j})$,
as shown in Fig.\ref{fig:links}(b) (and its Hermite conjugate).
If we take the non-compact formulation, the NLO effective action reads
\begin{equation}
\!\!\!\!\!\!\!\!
S_{\chi}^{(1) {\rm NC}} = -2\beta (-1)^n \sum_{x^{(3)}}\tilde{V}_4(x) 
 -\frac{\beta}{4}\sum_{x^{(3)}}\sum_{j=1,2}\tilde{V}_4(x)\tilde{V}_4(x+\hat{j}) +\mathrm{(const.)}
\label{eq:4-fermi-1-NC},
\end{equation}
where the fermionic bilinear $\tilde{V}_4$ is defined as
\begin{equation}
\tilde{V}_4(x) \equiv \bar{\chi}(x)\chi(x+\hat{4})-\bar{\chi}(x+\hat{4})\chi(x).
\end{equation}
Here we have used the integration formulae
\begin{eqnarray}
&& \frac{1}{2n\pi}\int_{-n\pi}^{n\pi} d\theta \; \theta = 0 \\
&& \frac{1}{2n\pi}\int_{-n\pi}^{n\pi} d\theta \; \theta e^{\pm i\theta} = \mp i(-1)^n \\
&& \frac{1}{2n\pi}\int_{-n\pi}^{n\pi} d\theta \; \theta^2 e^{\pm i\theta} = 2(-1)^n,
\end{eqnarray}
where $n$ is an arbitrary positive integer which specifies the interval of integration.
We will show afterwards that the ultimate result with mean field approximation
is independent of the choice of $n$.
It should be noted that the effect of the gauge field in the $x_3$ (or $z$)-direction
starts to appear from $O(\beta^3)$ in the strong coupling expansion.
(In $O(\beta^2)$, it only affects the constant term.)

\subsection{LO effective action with auxiliary field}

The non-local four-fermi interaction in  the LO effective action (Eq.(\ref{eq:4-fermi-0}))
can be linearized by the extended Stratonovich--Hubbard (ESH) transformation \cite{Miura:2009nu}:
\begin{equation}
 e^{\alpha AB} = \mathrm{(const.)} \times
 \int d\varphi d\varphi^* \exp[-\alpha(|\varphi|^2-A\varphi-B\varphi^*)],
\end{equation}
 where $A$ and $B$ are fermion bilinears and  $\alpha$ is a positive constant. 
Here we introduce a complex auxiliary field $\phi(x)$ corresponding to the LO term,
which transforms under the U(1)$_{_{\rm A}}$ rotation as
$\phi(x) \rightarrow e^{-2i \xi_{_{\rm A}} \epsilon(x)} \phi(x)$;
thus we separate the real and imaginary parts of $\phi(x)$ as
\begin{equation}
\phi(x) = \phi_\sigma(x) + i\epsilon(x)\phi_\pi(x).
\end{equation}
The real part $\phi_\sigma$ corresponds to the scalar operator $M=\bar{\chi}\chi$,
while the imaginary part $\phi_\pi$ to the pseudoscalar operator $P=\bar{\chi}i\epsilon\chi$.
With the auxiliary field $\phi$, we can rewrite Eq.(\ref{eq:4-fermi-0}) in terms of fermionic bilinears:
\begin{equation}
S_{\rm eff}^{(0)} = \frac{1}{4}\sum_{x^{(3)}}\left|\phi(x)\right|^2
-\sum_{x^{(3)},y^{(3)}} \bar{\chi}(x) \hat{V}[\phi](x,y) \chi(y), \label{eq:s-eff-0}
\end{equation}
where the matrix element $\hat{V}[\phi](x,y)$ is defined as
\begin{equation}
\hat{V}[\phi](x,y) = \frac{1}{4}\delta_{x,y}\left[\phi(x)+\phi^*(x-\hat{4})-4m_* \right] 
-\frac{1}{2}\sum_{j=1,2}\eta_j(x)\left[\delta_{y,x+\hat{j}}-\delta_{x,y+\hat{j}}\right].
\end{equation}
This matrix element is written in the momentum space as
\begin{equation}
\hat{V}[\phi](k,k') = \frac{1}{4}\left[\phi(p)+\phi^*(-p)e^{ip}- 4 m_* \delta_{p,0}\right]
 +\left[(i\sin k_1)\delta_{p,\pi \hat{4}} +(i\sin k_2)\delta_{p,\pi (\hat{4}+\hat{1})}\right],
\end{equation}
where we denote $p=k-k'$.
By integrating out the fermionic fields,
the effective action is written only in terms of $\phi$:
\begin{equation}
S_{\rm eff}^{(0)}[\phi] = \frac{1}{4}\sum_{x^{(3)}}\left|\phi(x)\right|^2 -\ln\det \hat{V}[\phi]. \label{eq:eff-phi}
\end{equation}

In order to diagonalize the matrix $\hat{V}[\phi]$ and calculate the determinant,
here we perform the mean-field approximation over $\phi_\sigma$ and $\phi_\pi$:
\begin{equation}
\phi(p) \rightarrow \phi_\sigma \delta_{p,0} +i\phi_\pi \delta_{p,\pi(\hat{4}+\hat{1}+\hat{2})}
\end{equation}
in the momentum space.
Thus the mass term in the effective action obtains the coefficient (``effective mass'')
$M_F = m_* -\phi_\sigma/2$, and $\hat{V}[\phi]$ yields the determinant
\begin{equation}
\ln\det \hat{V}[\phi] = \frac{N_S^2 N_T}{2}\int_\vect{k} \ln\left[G^{-1}(\vect{k};\phi)\right],
\end{equation}
with the effective bosonic propagator
\begin{equation}
G(\vect{k};\phi) = \left[\left|\frac{\phi}{2}-m_*\right|^2 +\sum_{j=1,2}\sin^2 k_j\right]^{-1} \label{eq:boson-prop}
\end{equation}
and the two-dimensional momentum integration
\begin{equation}
\int_\vect{k} = \int_{-\pi}^\pi \frac{dk_1}{2\pi} \int_{-\pi}^\pi \frac{dk_2}{2\pi}.
\end{equation}

\begin{figure}[tb]
\begin{center}
\includegraphics[width=8cm]{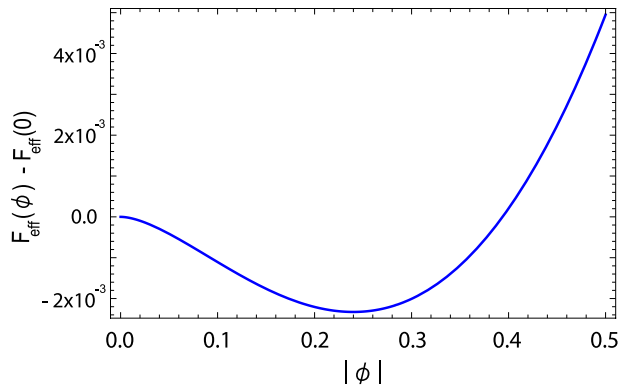}
\end{center}
\vspace{-0.5cm}
\caption{The free energy $F_\mathrm{eff}(\phi)$ in the lattice unit 
as a function of $|\phi|$ in the strong coupling limit ($\beta=0$) and 
 in the chiral limit ($m=0$).}
\label{fig:NLO}
\end{figure}

From the above arguments,
we can derive the free energy per unit cell (effective potential) analytically:
\begin{equation}
F_{\rm eff}^{(0)}(\phi) = \frac{1}{4}|\phi|^2 -\frac{1}{2}\int_\vect{k} \ln\left[G^{-1}(\vect{k};\phi)\right]
\label{eq:effpot-LO}
\end{equation}
Fig.\ref{fig:NLO} illustrates the LO free energy $F_{\rm eff}^{(0)}$ in the chiral limit ($m=0$).
From Eq.(\ref{eq:effpot-LO}), we find that
    $F_{\rm eff}(\phi \rightarrow \infty) \sim |\phi|^2$ 
  due to the tree-level term, while
  $F_{\rm eff}(\phi\rightarrow 0) \sim {\rm const.} + |\phi|^2 \ln |\phi|^2$  
  due to the fermion one-loop term.
Therefore $F_{\rm eff}^{(0)}$ takes the well-known ``Mexican hat'' shape,
 and dynamical  chiral symmetry breaking always takes place in the 
  strong coupling limit with the symmetry breaking pattern, 
  U(1)$_{_{\rm V}}$$\times$U(1)$_{_{\rm A}}$ $\rightarrow$ U(1)$_{_{\rm V}}$.
The potential minimum gives the chiral condensate
\begin{equation}
|\langle\bar{\chi}\chi\rangle| \equiv \sigma =|\langle\phi_\sigma\rangle|=0.240
\end{equation}
in lattice unit.

\subsection{NLO effective action}

There are two ways to linearize the NLO effective action Eqs.(\ref{eq:4-fermi-1}) and (\ref{eq:4-fermi-1-NC}).
One of them is perturbative expansion, which is rather straightforward;
we use the fermion propagator given by the inverse matrix of $\hat{V}[\phi]$:
\begin{eqnarray}
\left\langle \bar{\chi}(x)\chi(y) \right\rangle^{(0)} &=& \hat{V}^{-1}[\phi](x,y) \\
 &=& \delta_{x_4,y_4} \int_\vect{k} e^{i\vect{k}\cdot(\vect{x}-\vect{y})} G(\vect{k};\phi) \left[\frac{\phi^*}{2}-m_* + \sum_{j=1,2} \eta_j(y)(i\sin k_j)\right], \nonumber
\end{eqnarray}
where $\langle \cdots \rangle^{(0)}$ denotes the average over the LO effective action.
Thus we obtain
\begin{equation}
\langle V_j^{\pm}(x) \rangle^{(0)} = \pm \int_\vect{k} G(\vect{k};\phi) \sin^2 k_j,
\end{equation}
which enables us to calculate $\left\langle S_\chi^{(1) {\rm C}} \right\rangle^{(0)}$
and $\left\langle S_\chi^{(1) {\rm NC}} \right\rangle^{(0)}$
from Wick's theorem.
The NLO term of the free energy in the compact formulation reads
\begin{equation}
F_{\rm eff}^{(1){\rm C}}(\phi) = -\frac{\beta}{4} \sum_{j=1,2} \left[\int_\vect{k} G(\vect{k};\phi) \sin^2 k_j\right]^2. \label{eq:effpot-NLO}
\end{equation}
On the other hand, we find by this process that the only term in $S_\chi^{(1) {\rm NC}}$ remaining over the average is
\begin{equation}
\frac{\beta}{4} \sum_{x^{(3)}} \sum_{j=1,2}
\left[( V_j^+(x) V_j^-(x+\hat{4}) + (V_j^+ \leftrightarrow V_j^-) \right],
\end{equation}
and that all the other terms vanish up to $O(\beta^1)$.
Therefore the NLO effect in the non-compact formulation is twice larger than that in the compact formulation:
\begin{equation}
F_{\rm eff}^{(1){\rm NC}}(\phi) = 2F_{\rm eff}^{(1){\rm C}}(\phi). \label{eq:effpot-NLO-NC}
\end{equation}

Another way to obtain the same results is to introduce an auxiliary field,
which is rather indirect way than perturbative expansion,
but more convenient for further applications, such as observation of collective excitations.
By the ESH transformation, we introduce another complex auxiliary field $\lambda(x)$
corresponding to the NLO 4-fermi term $V^+V^-$.
Here we decompose $\lambda(x)$ into real and imaginary parts as $\lambda=\lambda_1+i\lambda_2$
and perform mean-field approximation over them.
The effective action in Eq.(\ref{eq:s-eff-0}) is modified by the (compact) NLO term as
\begin{equation}
S_\chi^{(0+1){\rm C}} = \frac{1}{4}\sum_{x^{(3)}}\left[\left|\phi\right|^2 + \frac{\beta}{4}\sum_{j=1,2}\left|\lambda\right|^2\right]
 -\sum_{x^{(3)},y^{(3)}} \bar{\chi}(x) \hat{V}[\phi,\lambda](x,y) \chi(y),
\end{equation}
where the modified matrix element $\hat{V}[\phi,\lambda]$ reads
\begin{eqnarray}
\hat{V}[\phi,\lambda](x,y) &=& \frac{1}{4}\delta_{x,y}\left[\phi(x)+\phi^*(x-\hat{4})-4m_* \right] \\
&& -\frac{1}{2} \left( 1 + \frac{\beta\lambda_2}{2} \right) \sum_{j=1,2} \eta_j(x) \left[e^{i\beta\lambda_1/2}\delta_{y,x+\hat{j}} - e^{-i\beta\lambda_1/2}\delta_{x,y+\hat{j}}\right]. \nonumber
\end{eqnarray}
Thus the effective boson propagator in Eq.(\ref{eq:boson-prop}) is replaced by
\begin{equation}
G(\vect{k};\phi,\lambda) = \left[ \left|\frac{\phi}{2} - m_* \right|^2 + \sum_{j=1,2} \left( 1 + \frac{\beta\lambda_2}{2} \right)^{2} \sin^2 \left( k_j - \frac{\beta\lambda_1}{2} \right) \right]^{-1} , \label{eq:boson-prop-NLO}
\end{equation}
and the effective potential in Eq.(\ref{eq:effpot-LO}) is modified correspondingly.

By requiring the stationary condition $\partial F_{\rm eff}^{(0+1){\rm C}}/\partial \lambda = 0$,
we can eliminate $\lambda$ by the relations
\begin{eqnarray}
\lambda_1 &=& -\int_\vect{k} G(\vect{k},\phi) \sin k_i \cos k_i + O(\beta) = 0+ O(\beta) \\
\lambda_2 &=& -\int_\vect{k} G(\vect{k},\phi) \sin^2 k_i + O(\beta),
\end{eqnarray}
leading to the effective potential in the same form as the sum of Eqs.(\ref{eq:effpot-LO}) and (\ref{eq:effpot-NLO}).
As for the non-compact gauge formulation,
the effective potential as the sum of Eqs.(\ref{eq:effpot-LO}) and (\ref{eq:effpot-NLO-NC}) can be derived by the same process as for the compact formulation.
Hereafter we mainly take the compact gauge formulation;
non-compact results can be easily derived
by substituting $\beta$ in compact results by $2\beta$ up to the NLO terms.

\begin{figure}[tb]
\begin{center}
\includegraphics[width=9cm]{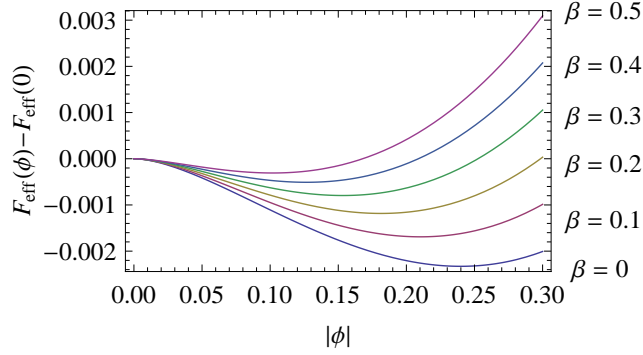}
\end{center}
\vspace{-1cm}
\caption{The free energy $F_{\rm eff}^{(0+1){\rm C}}(\phi)$ up to NLO terms with the compact gauge formulation in the chiral limit.}
\label{fig:NLO-2}
\end{figure}

\begin{figure}[!]
\begin{center}
\includegraphics[width=8cm]{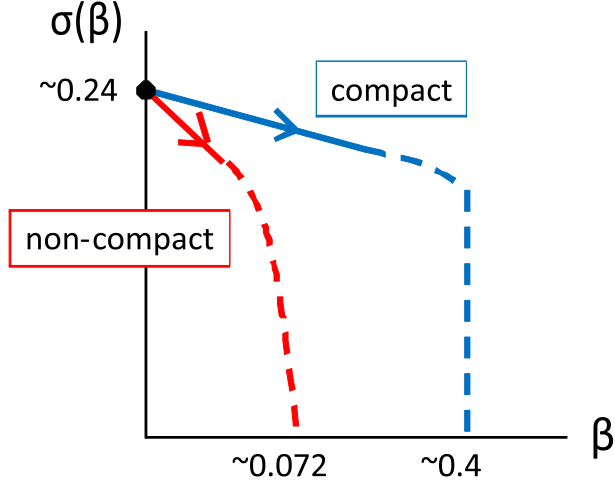}
\end{center}
\vspace{-0.5cm}
\caption{Schematic picture of the behavior of exciton condensate $\sigma(\beta)$ in the strong coupling region.
The compact (C) gauge formulation and the non-compact (NC) one coincides at $\beta=0$,
but $\sigma(\beta)$ with NC drops faster than that with C.
Behavior in the region far from $\beta=0$ (dashed curves) and the critical coupling value $\beta_C$ cannot be examined by strong coupling expansion,
so that here we employ the results suggested by lattice Monte Carlo method \cite{drut_2010}.}
\label{fig:conden}
\end{figure}

Since the NLO correction term $F_{\rm eff}^{(1){\rm C,NC}}(\phi)$ grows as $|\phi|$ increases,
the chiral condensate $\sigma$ is a decreasing function of $\beta$.
In other words, the chiral symmetry gets eventually restored as the coupling strength becomes weaker.
Up to the linear terms in $\beta$ and $m$, we can calculate $\sigma$ with the compact formulation as
\begin{equation}
\sigma^{{\rm C}}(\beta,m) \simeq (0.240 - 0.297 \beta + 0.0239\ ma) a^{-2}, \label{eq:sigma-exp}
\end{equation}
where we recover the lattice spacing $a$.
If we take $a^{-1}\simeq a^{-1}_{\rm Hc} = 1.39 \ {\rm keV}$ as a typical cutoff energy scale of this model,
we obtain $\sigma^{{\rm C}}(\beta,m) \simeq \left[\left(0.680 - 0.421 \beta + \frac{1.39\ m}{\rm eV} \right) {\rm keV} \right]^2$.
Since the NLO term with the non-compact formulation is twice that with the compact formulation,
the chiral condensate with the non-compact formulation $\sigma^{{\rm NC}}(\beta)$
drops twice faster than $\sigma^{{\rm C}}(\beta)$.
The behavior of $\sigma^{\rm C}(\beta)$ and $\sigma^{\rm NC}(\beta)$
around the strong coupling limit ($\beta=0$)
is schematically shown in Fig.\ref{fig:conden}.
The total fermion dynamical mass can be estimated from
the coefficient of the effective mass term,
\begin{equation}
M_F \equiv \frac{v_{_F}}{a}\frac{\sigma a^2}{2} +m.
\end{equation}
If we take $a \sim a_{_{\rm Hc}}$ and the compact formulation, we can estimate
$M_F \simeq (0.523 - 0.623 \beta ) \ {\rm eV} \ + 3.05 m$.

Since strong coupling expansion is valid only for small-$\beta$ region,
it is no longer possible to investigate the nature of the chiral (semimetal--insulator) transition:
the critical coupling value $\beta_C$, the order of the transition,
and the critical exponents for the transition,
unless some resummation techniques are introduced.
Monte Carlo simulations provide some clues to such critical behavior \cite{drut_2010}:
with the same lattice model employed in our studies,
the phase transition is estimated to be of second order
with $\beta_C=0.072 \pm 0.003$ with the non-compact formulation.
With the compact formulation,
it is estimated to be of first order with $\beta_C=0.42 \pm 0.01$.

Due to the finite-size effect of lattice,
lattice Monte Carlo simulations cannot reach the strong coupling and chiral limit.
Here we approximately compare our results with the Monte Carlo results
in the strong coupling and chiral limit,
by extrapolating the Monte Carlo results with the equation of state
which has been conjectured in analogy with the (3+1)-dimensional QED \cite{drut_2010}.
The chiral condensate reaches the value around $0.24$ (in lattice unit) in the strong coupling limit
by Monte Carlo simulation, both with the compact formulation and with the non-compact one.
By increasing the value of $\beta$,
it can be seen that $\sigma^{\rm NC}(\beta)$ drops faster than $\sigma^{\rm C}(\beta)$.
These results are consistent with the results obtained from our work.
We expect that our analytical studies, applicable around the strong coupling and chiral limit,
and the Monte Carlo simulation results, applicable within the finite coupling strength and finite mass region,
are complementary to each other.

\section{Collective Excitations}
\label{sec:col-ex}

\begin{figure}[htbp]
\begin{center}
\includegraphics[width=8.5cm]{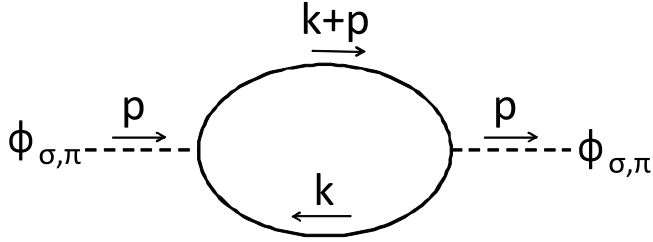}
\end{center}
\caption{The excitonic propagator $D_{\phi_{\sigma,\pi}}$ consists of a fermion one-loop diagram with two $\phi_{\sigma,\pi}$ legs,
with the overall momentum $\vect{p}$.}
\label{fig:feynmann}
\end{figure}

Here we consider two kinds of collective excitations,
or the fluctuations of the order parameter $\phi(x)$
around the symmetry broken state $\langle \phi \rangle = -\sigma$:
the phase fluctuation mode corresponding to $\phi_\pi(x)$
and the amplitude fluctuation mode corresponding to $\phi_\sigma(x)$,
which we name ``$\pi$-exciton'' and ``$\sigma$-exciton'' respectively
in analogy to the pion and the $\sigma$-meson in QCD.
Propagator of the $\alpha$-exciton mode ($\alpha=\sigma,\pi$) can be derived from
the effective action $S_{\rm eff}[\phi]$ (with the auxiliary field method) as
\begin{eqnarray}
D_{\phi_\alpha}^{-1}(x,y) &=&
 \left[\frac{\delta^2 S_{\rm eff}[\phi]}{\delta \phi_\alpha(x) \delta \phi_\alpha(y)}\right]_{\phi_\sigma=-\sigma,\phi_\pi=0}\\
&=& \frac{1}{2}\delta_{x_4,y_4} + {\rm Tr} \left[\hat{V}^{-1} \! \frac{\partial \hat{V}}{\partial \phi_\alpha(x)} \hat{V}^{-1} \frac{\partial \hat{V}}{\partial \phi_\alpha(y)}\right], \label{eq:D-prop1}
\end{eqnarray}
where we have used the formula
\begin{equation}
\frac{\partial^2 \ln \det \hat{V}}{\partial \phi_1 \partial \phi_2}
= -{\rm Tr} \left[ \hat{V}^{-1} \frac{\partial \hat{V}}{\partial \phi_1}\hat{V}^{-1} \frac{\partial \hat{V}}{\partial \phi_2} \right] + {\rm Tr} \left[ \hat{V}^{-1} \frac{\partial^2 \hat{V}}{\partial \phi_1 \partial \phi_2} \right] , \label{eq:formula-V}
\end{equation}
and the trace runs in the (2+1)-dimensional space-time.
The second term in Eq.(\ref{eq:formula-V}) vanishes
because the matrix element $\hat{V}[\phi](x,y)$ is linear in $\phi$.
The second term in Eq.(\ref{eq:D-prop1}) corresponds to the fermion one-loop diagram with two $\phi_\alpha$-legs,
as shown in Fig.\ref{fig:feynmann}.
Thus the propagators can be written in the momentum space as
\begin{equation}
{D}_{\phi_{\sigma,\pi}}^{-1}(\vect{p},i \omega_*) 
= \frac{1}{2}  - \frac{1+\cosh \omega_*}{8} \int_{\vect{k}}
H(\vect{k},\vect{p};\sigma) G(\vect{k};\sigma) G(\vect{k}+\vect{p};\sigma) ,\label{eq:correlation_1}
\end{equation}
where $G$ is the ``effective'' propagator defined in Eq.(\ref{eq:boson-prop}) (in LO)
and Eq.(\ref{eq:boson-prop-NLO}) (in NLO),
with $\phi$ and $\lambda$ substituted by their expectation values
determined by stationary conditions.
The effect of two vertices in the diagram is represented by
\begin{equation}
H(\vect{k},\vect{p};\sigma)
= \pm \left(m_*+\frac{\sigma}{2}\right)^{2} + \left(1+\frac{\beta\lambda_2}{2}\right)^{2} \sum_{j=1,2} \sin k_j \sin (k_j+p_j) ,
\end{equation}
where the $\pm$ sign corresponds to $\alpha=\pi$ and $\sigma$ respectively.
The dispersion relation for $\pi$ and $\sigma$
is derived from the pole of the propagator:
by restoring the temporal scaling by $v_{_F}$,
we obtain the dispersion relation
\begin{equation}
D_{\phi_{\sigma,\pi}}^{-1}\left(\vect{p},i\frac{\omega_{\sigma,\pi}(\vect{p})}{v_{_F}}\right)=0.
\end{equation}
Specifically the mass, or the excitation energy, of these modes
are given by the energy at zero momentum: $M_{\sigma,\pi}=\omega_{\sigma,\pi}(\vect{p}=0)$.

\subsection{$\pi$-exciton mode}

By considering the pole equation for $\pi$-exciton $D_{\phi_{\pi}}^{-1}(\vect{p}=0,\omega_*^{\pi})$,
the mass $M_\pi = v_{_F} \omega_*^{\pi}(\vect{p}=0)$ is given by
\begin{equation}
\cosh\omega_*^{\pi} = 4 \left[\int_{\vect{k}}G(\vect{k};\sigma)\right]^{-1} -1. \label{eq:omega-s}
\end{equation}
With the help of the gap equation
$\left[\partial F_{\rm eff}(\phi) / \partial \phi\right]_{\phi=-\sigma} =0$,
the integration part is given as
\begin{equation}
\int_{\vect{k}} G(\vect{k};\sigma) = 2 \frac{M_F(m=0)}{M_F(m)},
\end{equation}
where $M_F(m=0)$ is dynamically generated mass in the chiral limit
$M_F(m=0) = v_{_F}\sigma/2$.
Thus Eq.(\ref{eq:omega-s}) yields $\cosh\omega_*^{\pi} = 1+2m/M_F(m=0)$.
By taking the first order in the fermion bare mass $m$
and restoring the lattice unit $a$,
the mass $M_\pi(m)$ of this mode reads
\begin{eqnarray}
\label{eq:mpi}
 M_\pi \simeq \frac{2v_{_F}}{a} \sqrt{\frac{m}{M_F(m=0)}}
\end{eqnarray}
Thus the $\pi$-exciton serves as the massless Nambu--Goldstone (NG) boson
related to the spontaneous breaking of the ${\rm U(1)_A}$ symmetry,
which is emergent in the low-energy approximation.
As long as $0 \le m < 2 \ {\rm meV}$ is satisfied, 
$M_{\pi} < M_F $ holds, so that the $\pi$-exciton serves as the 
 lightest mode in  the system.
The $\pi$-exciton behavior derived here is reliable
as long as the excitation energy $M_\pi$ is within the scale of the low-energy approximation,
or, in other words, $m \ll M_F(m=0)$.

The relation $M_{\pi} \propto \sqrt{m}$ is similar to
the Gell-Mann--Oakes--Renner (GMOR) relation for the pion 
obtained from current algebra in QCD \cite{gmor,Hatsuda:1994pi}.
Considering the infinitesimal local ${\rm U(1)_A}$ transformation
\begin{equation}
\delta\chi(x) = i\epsilon(x)\alpha(x)\chi(x)
\end{equation}
and taking the transformation of $\langle P(x) \rangle$ in the present system,
we obtain 
\begin{eqnarray}
0= \delta \langle P(x) \rangle &=&  \frac{1}{Z} \int [d\bar{\chi}d\chi][d\theta] \delta\left[P(y) e^{-S}\right] \label{eq:WTI-1}\\
 &=& \left\langle \delta P(y)-P(y)\delta S \right\rangle, \nonumber
\end{eqnarray}
where the infinitesimal transformations of the pseudoscalar density and the action read
\begin{eqnarray}
\delta P(y) &=& -2\alpha(y) M(y) \\
\delta S &=& \sum_{x^{(3)}} \alpha(x) \left[-\partial_\mu J_\mu^{\rm axial}(x) +2m_* P(x)\right],
\end{eqnarray}
with the axial current defined as $J^{\rm axial}_{\mu}(x) \equiv \frac{i}{2} \epsilon(x) (V_{\mu}^-(x) - V_{\mu}^+(x))$.
By applying the functional derivative $\delta/\delta \alpha(x)$ to both sides of Eq.(\ref{eq:WTI-1}),
we obtain the axial Ward--Takahashi (WT) identity
\begin{equation}
\left \langle P(y) \left(\partial_\mu J_\mu^{\rm axial}(x) -2m_* P(x)\right) -2M(y)\delta_{xy} \right \rangle =0.
\end{equation}
Saturating this WT identity by the pole of the $\pi$-exciton, we obtain
\begin{equation}
\int d^3 p \langle 0 | P(y) | \pi(p) \rangle \frac{e^{-ipx}}{p^2} \langle \pi(p) | p_\mu J_\mu^{\rm axial}(x) +2m_* P(x) | 0 \rangle 
 = -2\sigma \delta_{xy} \label{eq:WTI-2}
\end{equation}
If we take $x=y=0$ and $m_*=0$, the matrix element
\begin{equation}
\langle 0 | P(0) | \pi(p) \rangle = -\frac{\sigma}{F_\pi^\tau} +O(m) \label{eq:WTI-3}
\end{equation}
is given in the leading order of $m$, where the temporal ``pion decay constant'' $F_{\pi}^{\tau}$ is defined
 by the matrix element, $\langle 0 | J^{\rm axial}_4(0)| \pi(p) \rangle = 2  F_{\pi}^{\tau} p_{\pi}^\tau$.
On the other hand, if we take $x=0\neq y$ and finite $m_*$, we obtain
\begin{equation}
2M_\pi^2 F_\pi^\tau +2m_* \langle 0 | P(0) | \pi(p) \rangle =0. \label{eq:WTI-4}
\end{equation}
Thus, from Eqs.(\ref{eq:WTI-3}) and (\ref{eq:WTI-4}), we obtain,
in the leading order of $m$,
 \begin{equation}
(F^{\tau}_{\pi}{M_\pi})^2 = m \sigma , 
\ \ F^{\tau}_{\pi}= \frac{\sigma a^2}{\sqrt{8v_{_F}}} a^{-1/2}, \label{eq:gmor-like}
\end{equation}
where $\sigma$ takes the value in the chiral limit ($m=0$),
and the lattice unit $a$ is again restored here. 
The first equation in Eq.(\ref{eq:gmor-like}),
which tells that the mass of $\pi$-exciton is proportional
to the square root of the fermion bare mass, is indeed the same form as the GMOR relation.

\subsection{$\sigma$-exciton mode}

\begin{figure}[tbhp]
\begin{center}
\includegraphics[width=9cm]{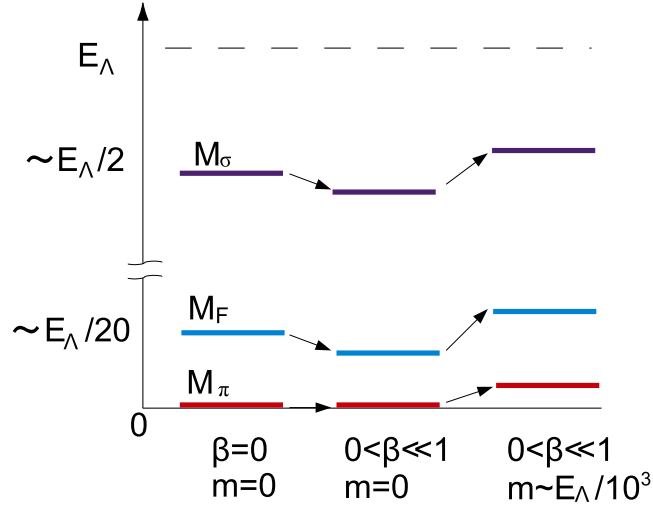}
\end{center}
\caption{A schematic picuture of the fermion excitation energy $M_F$,
 the $\pi$-exciton mass $M_{\pi}$ and 
 the $\sigma$-exciton mass $M_{\sigma}$ obtained 
 from the strong coupling expansion.
 $\beta=0$ and $m=0$ correspond to the strong-coupling limit and chiral limit,
 respectively.
 The ultraviolet cutoff for the energy, $E_\Lambda$, is given by $v_{_F} \pi/a$.}
\label{fig:spectrum}
\end{figure}

As for the $\sigma$-exciton, we obtain its mass
\begin{eqnarray}
M_{\sigma} &\simeq& (1.30 - 0.47 \beta)\frac{v_{_F}}{a}+22.6m
\end{eqnarray}
by solving 
 $D^{-1}_{\phi_{\sigma}}({\bf 0},iM_{\sigma}/v_{_F}) =0$ numerically.
This reduces to $M_\sigma \simeq (5.47 - 1.97 \beta)\  {\rm eV} + 22.6 m$
when we take $a \sim a_{_{\rm Hc}}$.
This value is comparable to the cutoff energy scale of the present lattice,
 $E_\Lambda = v_{_F} \pi/a = 13 \ {\rm eV}$.
Although the $\sigma$-exciton does not have
width in the present one-loop approximation, it would
eventually decay into a $\pi$-exciton pair in
higher orders. This is analogous to the situation for
the broad $\sigma$-meson in QCD \cite{Hatsuda:1994pi}.
However, since this value of $M_\sigma$ is beyond the scale of low-energy approximation,
its numerical accuracy is not reliable in the present approach.
In order to be more precise, the model without the low-energy approximation,
such as one preserving the structure of the original honeycomb lattice, is required.
Shown in Fig.\ref{fig:spectrum} is an illustration of the 
 spectrums of the fermion and collective exciations obtained in this study.

\section{Conclusion}
\label{sec:conclusion}

In this paper, we have investigated the behavior of the chiral symmetry of monolayer graphene
in the strong coupling regime analytically
by using the low-energy effective field (``reduced QED'') model of monolayer graphene.
A detailed explanation about the strong coupling expansion of the lattice-regularized model,
which was first applied for graphene model in our previous paper \cite{Araki_Hatsuda_2010}, is given.
``Chiral symmetry'' (sublattice on-site energy balance) of the original honeycomb lattice,
of the continuum effective theory, and of the staggered fermion formulation, are compared.
Since the compact U(1) gauge theory suffers from anomalous monopole condensation,
we compared the results with compact and non-compact formulations of the gauge field action.
As a result, whichever formulation we take, chiral symmetry is spontaneously broken in the strong coupling limit
with equal value of chiral condensate.
As the coupling strength becomes weaker, chiral condensate with the non-compact formulation drops faster
than that with the compact formulation.
These results up to NLO terms in strong coupling expansion agree with
the Monte Carlo simulation results, extrapolated to the strong coupling limit by the equation of state.

We have also estimated the energy scale of the emergent fermionic and collective (bosonic) excitations,
and have found that the fermionic quasiparticle and the $\pi$-exciton are within the scale of the low-energy approximation,
as long as the on-site energy difference between two sublattices is within that scale.
On the other hand, the mass of $\sigma$-exciton is quite large even in the chiral limit,
and it is beyond the low-energy approximation.
Thus one can say that the $\sigma$-exciton needs more investigation with a model without low-energy approximation,
such as one preserving the structure of the original honeycomb lattice.
Moreover, since the physical behavior of multilayer graphene,
in which interlayer hopping amplitude depends on the stacking pattern between layers, is also of great importance,
an analytic investigation with such a honeycomb lattice model is now strongly required.
We expect that the U(1) gauge theory on the honeycomb lattice can be suitable both for analytic calculations
and for numerical simulations.

Extension of this analysis to the finite temperature and finite density region is
of another important question.
In order to extend our analysis to the finite temperature region,
we have to take a temporally finite-size lattice
and incorporate fluctuation effects beyond the mean-field approximation.
This procedure would also enable us to
treat the dynamics in the temporal direction distinctively
so as to investigate the renormalization effect on the Fermi velocity.
Finite temperature and chemical potential analysis would be required not only for the chiral transition
but also for transport properties, such as electric conductivity and Hall effect.

\section*{Acknowledgements}
The author thanks H. Aoki, T. Hatsuda, T.Z. Nakano, Y. Nishida, 
 A. Ohnishi, T. Oka,  S. Sasaki, E. Shintani and  N. Yamamoto
  for discussions.
This work is supported by Grant-in-Aid for Japan Society for the Promotion of Science (DC1, No.22.8037).



\begin{thebibliography}{99}

 
\bibitem{wallace_1947} P.~E.~Wallace, Phys.~Rev.~\textbf{71}, 622 (1947).

\bibitem{Semenoff:1984dq}
  G.~W.~Semenoff,
  Phys.\ Rev.\ Lett.\  {\bf 53}, 2449 (1984).
  
 \bibitem{castro_neto_2009} See, e.g. A.~H.~Castro Neto et al.,
 Rev.~Mod.~Phys.~\textbf{81}, 109 (2009).

\bibitem{CN09} See, e.g., A.~H.~Castro Neto, Physics \textbf{2}, 30 (2009).

\bibitem{Hatsuda:1994pi}
See, e.g., K.~Fukushima and T.~Hatsuda,
 Rep.~Prog.~Phys.~ (2010), in press [arXiv:1005.4814 [hep-ph]].

\bibitem{Bolotin_2008}
K.~I.~Bolotin, K.~J.~Sikes, J.~Hone, H.~L.~Stormer, and P.~Kim,
Phys.~Rev.~Lett.~\textbf{101}, 096802 (2008).
Reviewed in,
V.~Crespi,
Physics \textbf{1}, 15 (2008).
 
 \bibitem{khveshchenko_2001} 
D.~V.~Khveshchenko, Phys.~Rev.~Lett.~\textbf{87}, 246802 (2001);
D.~V.~Khveshchenko and H.~Leal, Nucl.~Phys.~B \textbf{687}, 323 (2004);
D.~V.~Khveshchenko, J.~Phys.: Condens.~Matter \textbf{21}, 075303 (2009).

 \bibitem{gorbar_2002}
  E.~V.~Gorbar, V.~P.~Gusynin, V.~A.~Miransky and I.~A.~Shovkovy, Phys.~Rev.~B {\bf 66}, 045108 (2002).

 \bibitem{gorbar_2009}
  O.~V.~Gamayun, E.~V.~Gorbar and V.~P.~Gusynin, Phys.~Rev.~B {\bf 81}, 075429 (2010).
 
\bibitem{Giuliani_2010}
A.~Giuliani, V.~Mastropietro and M.~Porta,
arXiv:1001.5347 [cond-mat.str-el];
arXiv:1005.2528 [cond-mat.str-el].
 
\bibitem{Herbut_2006}
I.~F.~Herbut,
Phys.~Rev.~Lett.~\textbf{97}, 146401, (2006).

\bibitem{son_2007} D.~T.~Son, Phys.~Rev.~ B \textbf{75}, 235423 (2007).

\bibitem{son_2008} J.~E.~Drut and D.~T.~Son, Phys.~Rev.~B \textbf{77}, 075115 (2008).


 \bibitem{hands_2008} S.~Hands and C.~Strouthos, Phys.~Rev.~B \textbf{78}, 165423 (2008); W.~Armour, S.~Hands and C.~Strouthos, arXiv:0910.5646 [cond-mat.str-el].

 \bibitem{drut_2009} J.~E.~Drut and T.~A.~L\"{a}hde, Phys.~Rev.~Lett.~\textbf{102}, 026802 (2009); Phys.~Rev.~B \textbf{79}, 165425 (2009).

\bibitem{giedt_2009}
J.~Giedt, A.~Skinner and S.~Nayak, arXiv:0911.4316 [cond-mat.str-el].

\bibitem{drut_2010}
  J.~E.~Drut, T.~A.~L\"{a}hde and L.~Suoranta,
  arXiv:1002.1273 [cond-mat.str-el].

\bibitem{Drouffe:1983fv}
  Reviewed in J.~M.~Drouffe and J.~B.~Zuber,
  Phys.\ Rept.\  {\bf 102}, 1 (1983).

\bibitem{Nishida:2003uj}
  Y.~Nishida, K.~Fukushima and T.~Hatsuda,
  Phys.\ Rept.\  {\bf 398}, 281 (2004),  

\bibitem{Miura:2009nu}
  K.~Miura, T.~Z.~Nakano, A.~Ohnishi and N.~Kawamoto,
  Phys.\ Rev.\  D {\bf 80}, 074034 (2009).

\bibitem{Araki_Hatsuda_2010}
Y.~Araki and T.~Hatsuda,
Phys.~Rev.~B \textbf{82}, 121403(R) (2010).

\bibitem{gmor} M.~Gell-Mann, R.~J.~Oakes and B.~Renner, Phys.~Rev.~\textbf{175}, 2195 (1968).

\bibitem{gorbar_2001}
E.~V.~Gorbar, V.~P.~Gusynin, V.~A.~Miransky,
Phys.~Rev.~D \textbf{64}, 105028 (2001).

\bibitem{reich} S.~Reich, J.~Maultzsch, C.~Thomsen and P.~Ordej\'{o}n, Phys.~Rev.~B \textbf{66}, 035412 (2002).

 \bibitem{zhou_2007} S.~Y.~Zhou et al.,
Nature Materials \textbf{6}, 770 (2007).
 
\bibitem{bai_2010} See e.g., J. W. Bai et al.,
Nature Nanotechnology \textbf{5}, 190 (2010).

\bibitem{Nielsen-Ninomiya}
H.~B.~Nielsen and M.~Ninomiya,
Nucl.~Phys.~B \textbf{185}, 20 (1981);
erratum B \textbf{195}, 541 (1981);
Nucl.~Phys.~B \textbf{193}, 173 (1981).

\bibitem{Susskind_1977}
L.~Susskind,
Phys.~Rev.~D \textbf{16}, 3031 (1977).

\bibitem{Sharatchandra_1981}
H.~S.~Sharatchandra, H.~J.~Thun and P.~Weisz,
Nucl.~Phys.~B \textbf{192}, 205 (1981).

\bibitem{monopole-cond}
J.~B.~Kogut and E.~Dagotto,
Phys.~Rev.~Lett.~\textbf{59}, 617 (1987);
J.~B.~Kogut and C.~G.~Strouthos,
Phys.~Rev.~D \textbf{67}, 034504 (2003).

\bibitem{Kogut:1982ds}
  J.~B.~Kogut,
  Rev.\ Mod.\ Phys.\  {\bf 55}, 775 (1983).

\bibitem{Wilson_1974}
K.~G.~Wilson,
Phys.~Rev.~D \textbf{10}, 2445 (1974).

\bibitem{Kawamoto_Smit_1981}
N.~Kawamoto and J.~Smit, Nucl.~Phys.~B \textbf{192}, 100 (1981).

\bibitem{Nishida_2004}
Y.~Nishida,
Phys.~Rev.~D \textbf{69}, 094501 (2004).

\bibitem{Kawamoto_2007}
N.~Kawamoto, K.~Miura, A.~Ohnishi, and T.~Ohnuma,
Phys.~Rev.~D \textbf{75}, 014502 (2007).

\bibitem{Ohnishi_2007}
A.~Ohnishi, N.~Kawamoto, K.~Miura, K.~Tsubakihara, and H.~Maekawa,
Prog.~Theor.~Phys.~Suppl.~\textbf{168}, 261 (2007).

\bibitem{TZN_2010_1}
T.~Z.~Nakano, K.~Miura and A.~Ohnishi,
Prog.~Theor.~Phys.~\textbf{123}, 825 (2010).

\bibitem{TZN_2010_2}
T.~Z.~Nakano, K.~Miura and A.~Ohnishi,
arXiv:1009.1518 [hep-lat].



\end{thebibliography}
\end{document}